# Surfactant assisted synthesis of Co and Li doped ZnO nanocrystalline samples showing room temperature ferromagnetism

By Onattu D. Jayakumar, Iyyani K. Gopalakrishnan* and Shailendra K. Kulshreshtha

The dilute magnetic semiconductor oxides (DMSO) are of current interest because of their potential 'spintronics' applications, where the charge and spin degrees of freedom of electrons are used simultaneously for novel memory and optical device applications.[1,2] In particular, Co doped ZnO has attracted considerable interest.[3] There have been a number of reports about the observance of room temperature ferromagnetism in thin films of $Zn_{1-x}Co_xO$ produced by different techniques.[4-7] Schwartz et al.[8] also observed ferromagnetism above room temperature in aggregated particles of Co doped ZnO, by heat treating (below 200°C) colloidal quantum dots of Co doped ZnO. However, most recent works on well-characterized polycrystalline $Zn_{1-x}Co_xO$ samples indicate that they are not ferromagnetic at room temperature,[9-14] except for an isolated report by Deka et al..[15] In general, studies on polycrystalline samples have converged on to a conclusion that robust room temperature ferromagnetism (RTF) is not realizable in Co doped ZnO without additional carrier doping. Sato and Katayama-Yoshida[16] predicted Co doped ZnO would become ferromagnetic in the presence of n-type carriers. This was experimentally demonstrated by Schwartz and Gamelin.[17] They were the first to show the reversible cycling of paramagnetic (P) to ferromagnetic (FM) state in Co doped ZnO spin coated films, produced from colloidal nanocrystals, by introducing and removing interstitial Zn ($Zn_i$), a native n-type defect of ZnO. Later Spaldine,[18] in a computational study, showed only hole doping promotes RTF in Co doped ZnO. This is in contrast with RTF observed in electron doped Co doped ZnO by Schwartz and Gamelin.[17] Through an analysis of density functional calculations Sluiter et al.[19] have shown that both electron doping with zinc interstitials and hole doping with zinc vacancies make ZnO:Co strongly FM. Sluiter et al.[19] also showed experimentally that co-doping with Li promotes FM in Co doped ZnO. Here they assumed Li co-doping creates p-type carriers in ZnO. Bergqvist et al.,[20] combining first principles calculations of interatomic exchange interactions with a Heisenberg model and Monte Carlo simulations, have explained the preparation dependence of ferromagnetism in transition metal doped dilute magnetic semiconductors as due to the random ordering of magnetic ions and distribution of defects. Hong et al.[21] in a recent paper highlighted the role of defects in tuning ferromagnetism in DMSO films. In view of this, we thought it is worth while to explore new ways to fine tune synthesis conditions to optimize magnetic characteristics of Co doped ZnO polycrystalline samples. In this communication, we present our efforts in fine tuning preparation conditions for optimizing ferromagnetic characteristics of Co doped ZnO co-doped with Li. Recent neutron scattering study of polycrystalline samples of $Zn_{1-x}Co_xO$ (x= 0, 0.05, 0.10, 0.20, 0.30), by Lee et al.[22] have shown that they are monophasic only up to x = 0.05. In view of this, we have chosen Co doping concentration to 5 at %. We have shown that high-$T_c$ ferromagnetism in Co and Li doped ZnO nanocrystalline samples can be activated by surfactant treatment. The new synthesis method described in this communication is of particular interest as it is not only reproducible but also found to enhance the magnetic moment.

Rietveld profile refinement analysis of XRD data of $Zn_{0.95}Co_{0.05}O$ and $Zn_{0.85}Co_{0.05}Li_{0.10}O$ prepared by the surfactant assisted synthesis provided convincing evidence that they are single phase with the wurtzite structure (space group $P6_3mc$) (Fig.1). Hexagonal cell parameters *a* and *c* of $Zn_{0.85}Co_{0.05}Li_{0.10}O$ extracted by Rietveld analysis are 3.2630(2) and 5.2179(1) Å respectively while that of $Zn_{0.95}Co_{0.05}O$ are 3.2559(1) and 5.2130(2) Å. Inset in Fig.1 shows the XRD patterns of $Zn_{0.85}Co_{0.05}Li_{0.10}O$ with and without AOT treatment. It can be seen from the inset that width of peaks decreases considerably on surfactant treatment. The average crystallite size was determined from the extra broadening of the X-ray diffraction peaks of the sample using Scherrer`s formula applied to the strongest peak. The average crystallite size of as prepared samples is ~10 nm. The average crystallite sizes of Co doped ZnO with and without Li co-doping increases to 28 and 31 nm respectively after surfactant treatment.

*DC* magnetization loops of $Zn_{0.95}Co_{0.05}O$ samples (annealed at 675 K for 2 hours) with and without AOT (sodium bis (2-ethylhexyl) sulfosuccinate) treatment, recorded at RT, are presented in Fig.2. In addition, data on pristine ZnO treated with AOT is also presented. It can be seen that for the sample without AOT treatment the magnetization is linear indicating its paramagnetic nature while that treated with AOT is *S* shaped typical of ferromagnetic materials. The saturation magnetization ($M_s$) is 0.058 emu/g (0.0191$\mu_B$/Co). The pristine ZnO treated with AOT showed diamagnetic behaviour as depicted in Fig.2. Similar magnetic behaviour was observed when the same samples were treated with triblock copolymer (pluronic P123) (poly(ethylene glycol)-block-poly(propylene glycol)-block-poly(ethylene glycol)). This indicates that the observed ferromagnetism in surfactant treated doped samples is not coming from any impurities associated with surfactants. One can also preclude Co metal clusters as the source of RTF since the synthesis is carried out under aerobic conditions. In Fig.3, we have presented data on Co doped ZnO co-doped with Li samples of nominal composition $Zn_{0.85}Co_{0.05}Li_{0.10}O$ treated with two different surfactants viz. AOT and triblock copolymer (Pluronic P123). In addition, data on as synthesized sample is also shown for comparison. The inset in Fig.3 shows an expanded view of *M-H* loops around origin. It can be seen from Fig.3 that for the untreated sample, the *M-H* loop is linear with field indicating its paramagnetic nature, while that treated with AOT and copolymer are ferromagnetic. There is five times increase in the magnetization value for AOT treated sample ($M_s$ = 0.0479 $\mu_B/Co^{2+}$) as compared to the untreated one ($M_s$ = 0.0089 $\mu_B/Co^{2+}$). The increase of saturation magnetization value in the case of copolymer treated sample ($M_s$ = 0.0275 $\mu_B/Co^{2+}$) is only three times. The surfactant treated samples show a coercive field ($H_c$) of 116 ± 10 Oe. The value of $M_s$ observed for AOT treated sample is in agreement with the value reported for spin coated thin films of Co doped ZnO injected with n-type carriers by chemical manipulation.[23] It can be seen that the value of $M_s$ of $Zn_{0.95}Co_{0.05}O$ is 0.0191 $\mu_B/Co^{2+}$ while that of $Zn_{0.85}Co_{0.05}Li_{0.10}O$ is 0.0479 $\mu_B/Co^{2+}$. This shows that there is substantial increase in the $M_s$ value on co-doping it with lithium.

The appearance of ferromagnetism upon treatment of surfactant can be attributed to the fusion of defects at interfaces between nanocrystals and the randomization of dopant ions. Similar results was observed by Gamelin`s group.[8,17,23,24] Their work unambiguously demonstrated the role of interfacial or grain-boundary defects in activating room temperature ferromagnetism in a broad spectrum of transition metal doped oxides including cobalt-doped ZnO. It is likely that such lattice defects are more widely influential than currently recognized. The contrasting observations often reported for the same materials prepared by different laboratories or by different methods can be understood if one takes into consideration the role played by these defects in activating FM. It is worth mentioning that the crystallite size of $Zn_{0.85}Co_{0.05}Li_{0.10}O$ increases from 10 nm to 28 nm on treatment with surfactant. Bryan et al.[24] studying Co doped ZnO nanocrytals using electronic absorption spectroscopy as a dopant-specific in-situ spectroscopic probe, found $Co^{2+}$ ions are to be quantitatively excluded from the ZnO critical nuclei but incorporated nearly statistically in the subsequent growth layers, resulting in crystallites with pure ZnO cores and $Zn_{1-x}Co_xO$ shells. During the aggregation of crystallite size it is reasonable to speculate that the doped magnetic ions gets randomized in the host matrix in view of the literature reports that the presence of surfactant reduces interfacial tension between the host and the dopant.[25,26] This probably inhibits the clustering of the dopant ions facilitating the randomization of Co atoms in the ZnO lattice. This is schematically illustrated in Fig.4. Our findings are in agreement with the experimental work of Gamelin and coworkers[8,17,23,24] and are also in line with the computational study of Bergquist et al.[20] that the random distribution of magnetic ion and defects in the host lattice plays a crucial role on the appearance of ferromagnetism with high $T_c$ in dilute magnetic semiconductors. The enhanced ferromagnetic characteristics displayed by the Li co-doped samples indicate that the injection of carriers is required in addition to the randomization of magnetic ions and defects. This is in agreement with the studies of Sluiter et al.[19] and Kittilstved et al.[23] on Co doped ZnO. From first-principles density-functional calculations Lee and Chang [27] have shown that substitutional Li and Na are better acceptors in ZnO with shallow acceptor levels. However, Local density functional calculations by Wardle et al.[28] have shown that interstitial Li is more stable than substitutional Li in ZnO. Electron paramagnetic resonance and electron nuclear double resonance (ENDOR) experiments on Li/Na doped ZnO nanoparticles by Orlinskii et al.[29] observed the presence of shallow donors that are related to interstitial Li and Na atoms thus supporting the theoretical predictions of Wardle et al.[28] that Li and Na can enter the ZnO lattice interstitially and can act as shallow donors. This shows Li co-doping generates donors rather than acceptors. This explains the apparent contradictions between the works of Sluiter et al.[19] and Schwartz and Gamelin[17] on Li co-doped and Zn metal co-doped cobalt doped ZnO respectively.

The enhancement of $M_s$ value, on co-doping with Li indicates the observed ferromagnetism cannot be attributed to any impurity phases involving cobalt oxides or cobalt metal clusters. Our results also can explain the preparation dependent ferromagnetism observed by different groups in Zn-Co-O system. This synthetic procedure allows the relatively rapid preparation of gram quantities of Co doped ZnO in open containers with no need for any special precautions to avoid atmospheric contamination, or temperature stabilization. In view of this we feel this surfactant assisted synthesis method developed by us will come in a long way in resolving many ambiguities surrounding the origin of ferromagnetism in the Zn-Co-O system. This method of synthesis can also be extended to other transition metal doped ZnO, especially to Mn doped ZnO. Nevertheless, the $M_s$ value is far below the expected value of $3\mu_B$/Co atom for tetrahedral coordinated high spin $Co^{2+}$-ion. Further studies are in progress to find out the exact cause for the activation of ferromagnetism and also to further improve the magnetic characteristics of Co doped ZnO.

In summary, we have developed a simple, surfactant assisted synthesis route for the preparation, in gram quantities, of Co and Li doped ZnO nanocrystalline samples showing robust room temperature ferromagnetism. Our studies show that RTF is intrinsic to $Zn_{0.85}Co_{0.05}Li_{0.10}O$ and not due to any segregated secondary phases. In addition, it has been shown that the defects play an important role in activating RTF in these oxide systems. This also provide an explanation for the widely varying results observed in the literature. The method can also be extended for the synthesis of other transition metal doped ZnO.

Experimental

In this method zinc acetate di hydrate (99.99%), cobalt acetate tetra hydrate (99.99%), lithium acetate (99.99%), sodium bicarbonate ($NaHCO_3$) and absolute ethanol were used without any further purification. In a typical synthesis to prepare 5 at % Co doped ZnO samples the procedure is as follows. Zinc acetate di hydrate, Cobalt acetate tetra hydrate and sodium bi carbonate taken in appropriate proportion are mixed at room temperature. The mixture is pyrolysed[30] at 450 K for 2 hours and the product is then washed several times with de-ionized water and absolute ethanol to remove the sodium acetate and other impurities, and subsequently oven dried at 400 K for overnight. These samples were then mixed with surfactant AOT (sodium bis (2-ethylhexyl) sulfosuccinate) (99.9%) or triblock copolymer (pluronic P123 purity 99.9%) (poly (ethylene glycol)-block poly (propylene glycol)-block poly (ethylene glycol)) (sample to surfactant ratio 1:10) and heated at 675 K for 1 hour and subsequently washed and dried followed by calcining at 675 K for 2 hours in air. Same procedure has been followed to prepare Co (5 at %) doped Li (10 at %) co-doped ZnO samples. Phase purity and the structure of the samples were analyzed using CuKα radiation by employing a Philips Diffractometer (model PW 1071) fitted with graphite crystal monochromater. The lattice parameters of the compounds were extracted by Rietveld refinement of the XRD data by using the computer code Fullprof [31] with the X-ray intensity collected for the range $10° \leq 2\theta \leq 70°$. *DC* magnetization measurements as a function of field were carried out using an E.G.&G P.A.R vibrating sample magnetometer (model 4500).

[*] Dr. I. K. Gopalakrishnan, O. D. Jayakumar, Dr. S. K. Kulshreshtha

Chemistry Division,

Bhabha Atomic Research Centre,

Mumbai 400085, India

E-mail: ikgopal@magnum.barc.ernet.in

Figure captions

Fig.1 Rietveld refinement of Room temperature XRD data of $Zn_{0.95}Co_{0.05}O$ and $Zn_{0.85}Co_{0.05}Li_{0.10}O$ samples treated with AOT and copolymer. The dots represent the observed data while solid line through dots is the calculated profile. The vertical tics below the profile are the expected reflections for wurtzite phase. The difference pattern is given below the vertical tics. Inset shows the XRD patterns of $Zn_{0.85}Co_{0.05}Li_{0.10}O$ before and after AOT treatment.

Fig.2 Isothermal magnetization of $Zn_{0.95}Co_{0.05}O$ sample with and without AOT treatment and recorded at room temperature. Data on pristine ZnO treated with AOT is also shown.

Fig.3 Isothermal magnetization of $Zn_{0.85}Co_{0.05}Li_{0.10}O$ with and without surfactant treatment recorded at RT. Inset shows its expanded view at the origin.

Fig.4 Schematic illustration of the effect of surfactant treatment on the magnetic state of $Zn_{0.85}Co_{0.05}Li_{0.10}O$.

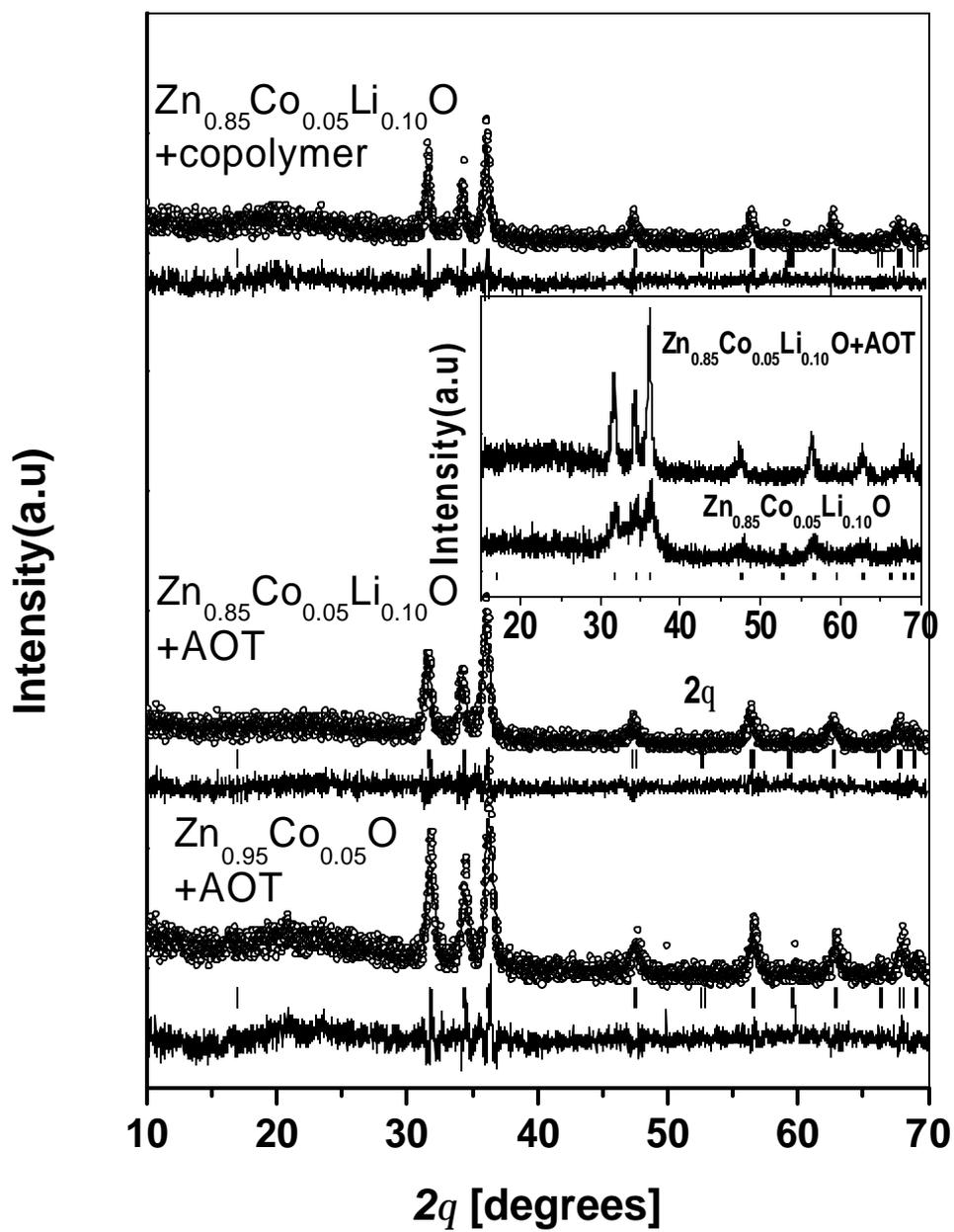

Fig.1

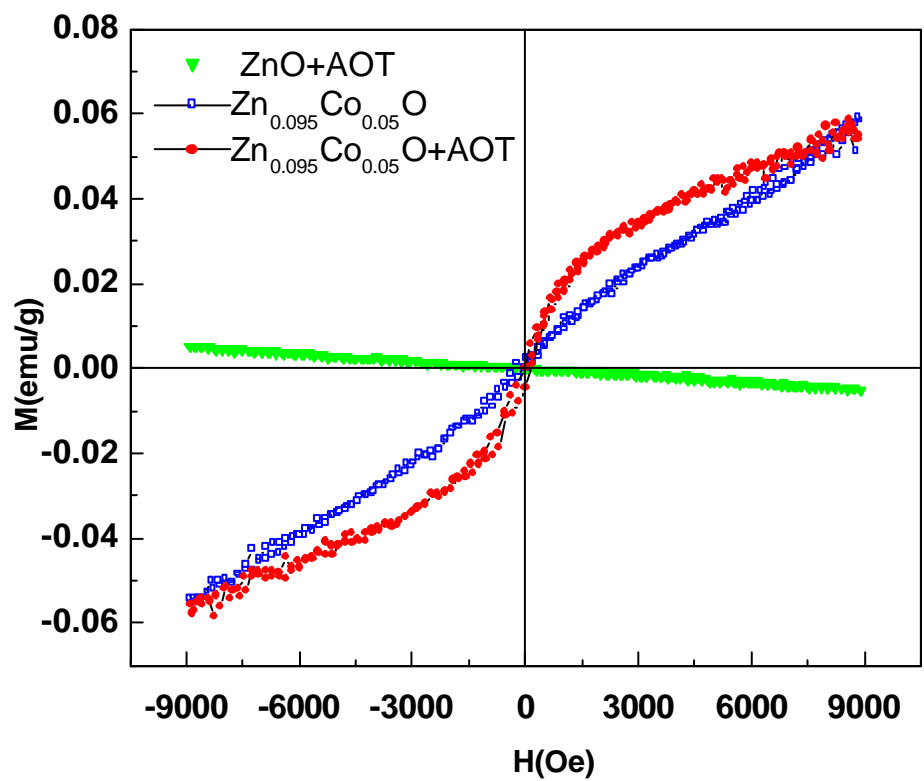

Fig.2

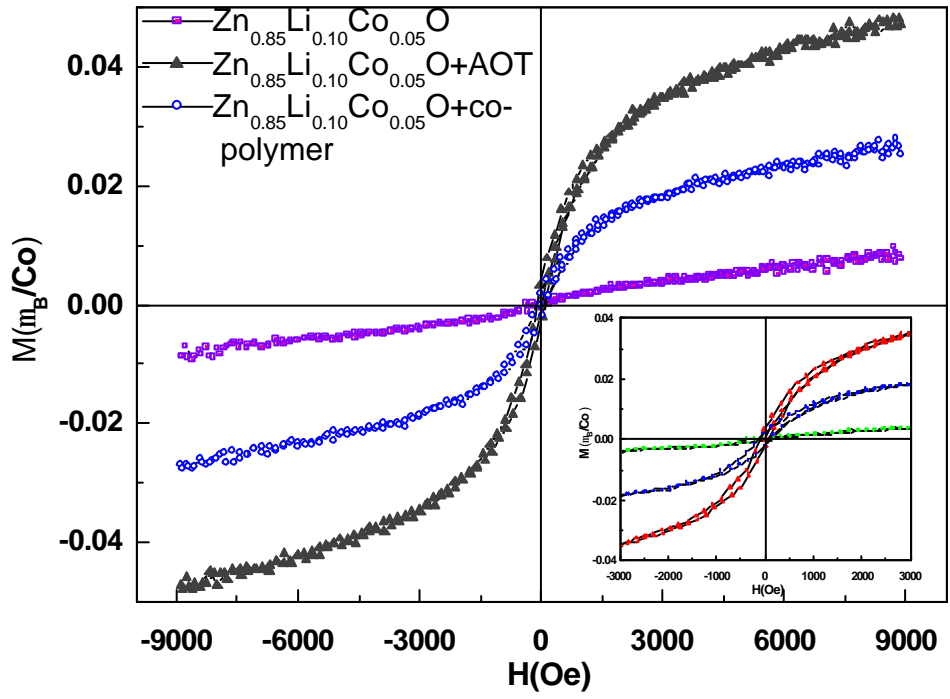

Fig.3

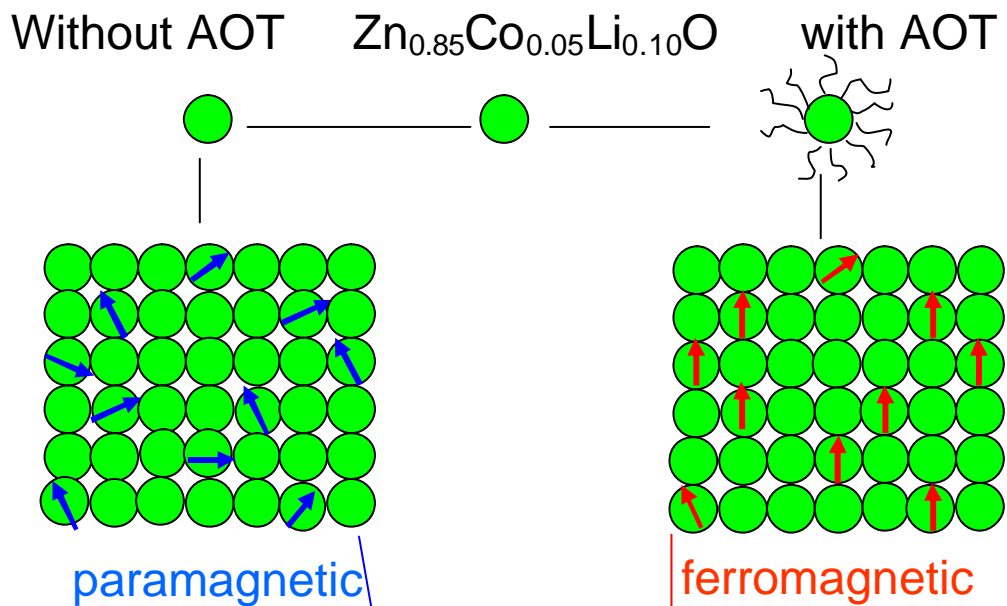

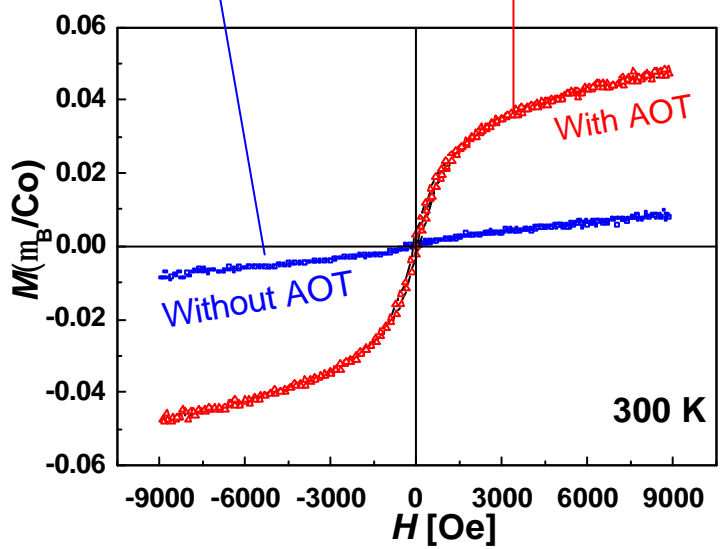

Fig.4